\documentclass[a4paper]{jpconf}
\usepackage{graphicx}
\usepackage{float}
\usepackage[square,sort&compress]{}
\usepackage{wrapfig, blindtext}
\usepackage{cite}
\usepackage{xcolor}

\begin{document}
\title{Photo-processing of astro-PAHs}

\author{C Joblin\textsuperscript{1}, G Wenzel\textsuperscript{1}, S Rodriguez Castillo\textsuperscript{1,2}, A Simon\textsuperscript{2}, H Sabbah\textsuperscript{1,3}, A Bonnamy\textsuperscript{1}, D Toublanc\textsuperscript{1}, G Mulas\textsuperscript{1,4}, M Ji\textsuperscript{1},  A~Giuliani\textsuperscript{5,6}, L Nahon\textsuperscript{5}}

\address{\textsuperscript{1}Institut de Recherche en Astrophysique et Plan\'{e}tologie (IRAP), Universit\'{e} de Toulouse (UPS), CNRS, CNES, 9 Avenue du Colonel Roche, F-31028 Toulouse Cedex 4, France}
\address{\textsuperscript{2}Laboratoire de Chimie et Physique Quantiques (LCPQ/IRSAMC), Universit\'{e} de Toulouse and CNRS,UT3-Paul Sabatier, 118 Route de Narbonne, 31062 Toulouse, France}
\address{\textsuperscript{3}Laboratoire Collisions Agr\'{e}gats R\'{e}activit\'{e} (LCAR/IRSAMC), Universit\'{e} de Toulouse (UPS), CNRS, 118 Route de Narbonne, F-31062 Toulouse, France}
\address{\textsuperscript{4}Istituto Nazionale di Astrofisica (INAF), Osservatorio Astronomico di Cagliari, Via della Scienza 5, I-09047 Selargius (Cagliari), Italy}
\address{\textsuperscript{5}Synchrotron SOLEIL, L'Orme des Merisiers, F-91192 Gif-sur-Yvette Cedex, France}
\address{\textsuperscript{6}INRA, UAR1008, Caract\'{e}risation et Elaboration des Produits Issus de l'Agriculture, F-44316 Nantes, France}

\ead{christine.joblin@irap.omp.eu}

\begin{abstract}
Polycyclic aromatic hydrocarbons (PAHs) are key species in astrophysical environments in which vacuum ultraviolet (VUV) photons are present, such as star-forming regions. The interaction with these VUV photons governs the physical and chemical evolution of PAHs. Models show that only large species can survive. However, the actual molecular properties of large PAHs are poorly characterized and the ones included in models are only an extrapolation of the properties of small and medium-sized species. We discuss here experiments performed on trapped ions including some at the SOLEIL VUV beam line DESIRS. We focus on the case of the large dicoronylene cation, C$_{48}$H$_{20}^+$, and compare its behavior under VUV processing with that of smaller species. We suggest that C$_2$H$_2$ is not a relevant channel in the fragmentation of large PAHs. Ionization is found to largely dominate fragmentation. In addition, we report evidence for a hydrogen dissociation channel through excited electronic states. Although this channel is minor, it is already effective below $13.6\,\mathrm{eV}$ and can significantly influence the stability of astro-PAHs. We emphasize that the competition between ionization and dissociation in large PAHs should be further evaluated for their use in astrophysical models.
\end{abstract}

\section{Introduction}
Polycyclic aromatic hydrocarbons (PAHs) are present in photodissociation regions (PDRs) associated with massive star-forming regions, such as the prototypical Orion Bar region \cite{Peeters2004}. Their interaction with vacuum ultraviolet (VUV) photons can trigger various molecular processes: -(i)- ionization resulting in gas heating by thermalisation of the emitted electrons \cite{Bakes1994}, -(ii)- photodissociation limiting the survival of PAHs and producing molecules such as H$_2$ and C$_2$H$_2$ in PDRs \cite{Allain1996, LePage2001,Montillaud2013, Andrews2016, Castellanos2018_PDR}, and -(iii)- radiative cooling leading to the well-known aromatic infrared (IR) emission bands between $3$ and $15\,\mathrm{\mu m}$, which constitute the only direct diagnosis we have so far for the presence of these large molecules in astrophysical environments as proposed in the initial PAH model \cite{Leger1984, Allamandola1985}.

The chemical evolution of PAHs in PDRs has been modelled by several authors \cite{Allain1996, LePage2001, Visser2007, Montillaud2013, Andrews2016}. These models determined a critical molecular size, typically of $50-60$ carbon atoms, below which these PAHs are not expected to survive. In some models, the critical size is estimated from the ability of the molecule to lose C$_2$H$_2$ since in PDRs the absorption of UV photons is faster than the chemistry that could rebuild the carbon skeleton. The PAH experiencing C$_2$H$_2$ loss will therefore be ultimately destroyed if its fragments also experience C$_2$H$_2$ loss. Other models focus on the more common hydrogen loss. In this case the competition with rehydrogenation by reactivity with the abundant H and H$_2$ species needs to be considered (see for instance \cite{Montillaud2013}).

All these chemical models rely on photophysical and chemical rates that are still only partly known for small and medium-sized species up to 24 carbon atoms. These species are easier to handle in gas-phase in the laboratory compared to large ones. However, their properties might differ, especially when one considers the interaction with VUV photons. In addition, one should keep in mind the extreme isolation conditions of astrophysical environments in which radiative cooling can play a key role in the photophysics of PAHs. The challenge for laboratory experiments is therefore to address the multiscale aspects of the photophysics of PAHs from very fast femtosecond processes such as ionization \cite{Marciniak2015} to the very long timescale of IR cooling (see Fig.\ref{cascade}). This challenge has to be addressed for large PAH species of $\sim\!50-100$ carbon atoms, which are the best candidates to survive in PDRs. This article reports our methodology and first results towards the study of large PAH species and evidence for their specificity.

\section{The photophysics of astro-PAHs}

\begin{wrapfigure}[24]{r}{0.5\linewidth}
\centering
\vspace*{-1cm}
\includegraphics[width=0.95\linewidth]{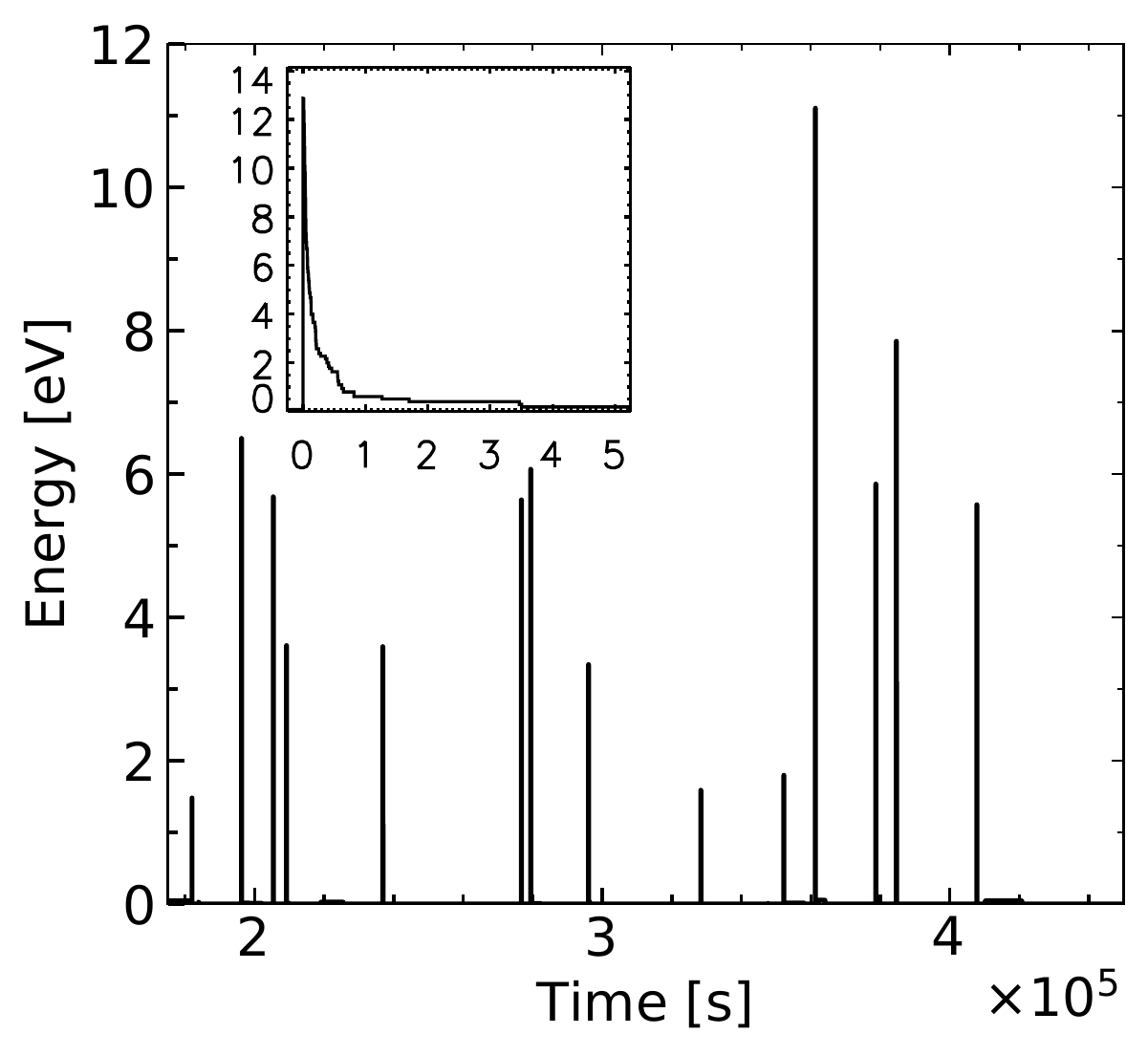}
\vspace{-0.5cm}
\caption{\label{cascade} Timescale for the absorption of UV photons by a coronene cation C$_{24}$H$_{12}^+$ in the NGC 7023 NW PDR studied in the chemical model of Montillaud et al. \cite{Montillaud2013}. An example of an IR cooling cascade is also provided in the inset, zoomed in over a 5~s time interval.
The calculations were performed using a dedicated Monte Carlo code to describe the photophysics of PAHs \cite{Mulas2006}.}
\end{wrapfigure}

In PDRs, astro-PAHs are well isolated. Infrared emission, which is a slow process and can extend over seconds (cf. inset graph in Fig. \ref{cascade}), is therefore a major process in energy relaxation due to the lack of collisions that could enter into competition. Considering a typical gas (mainly hydrogen) density of $\sim\!10^4\,\mathrm{cm^{-3}}$ in PDRs, the timescale for collisions is typically 28 hours taking an optimistic rate of collisions of $10^{-9}\,\mathrm{cm^3s^{-1}}$. The timescale for the absorption of a VUV photon depends on the photon flux. It is several hours in the NGC~7023 PDR, which was studied by Montillaud \textit{et al.} \cite{Montillaud2013} (cf. Fig. \ref{cascade}), and tens of minutes in the brighter Orion Bar. This leaves time to lose most of the internal energy by infrared cooling. The energy of the available VUV photons is usually below $13.6\,\mathrm{eV}$ due to the ionization of hydrogen atoms but can be higher ($\sim\!20\,\mathrm{eV}$) in the ionized bubbles around massive stars \cite{compiegne2007}.  

\begin{wrapfigure}[19]{r}{0.5\linewidth}
\vspace*{-0.5cm}
\centering
\includegraphics[width=\linewidth]{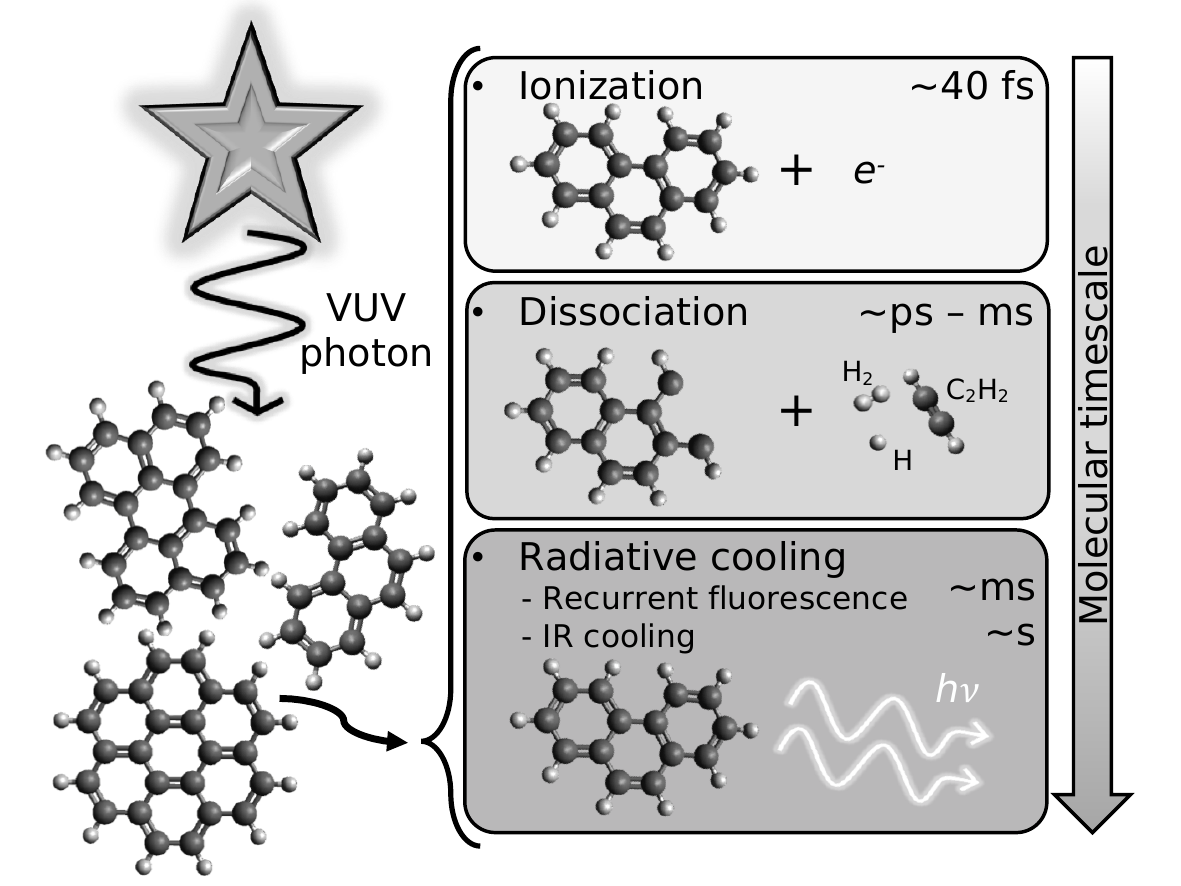}
\caption{\label{astropah_schematic}Schematic of the different relaxation processes of astro-PAHs upon VUV photon absorption and involved characteristic molecular timescales.}
\end{wrapfigure}

Since the initial description of the photophysics of an astro-PAH \cite{Leger1989} we can now draw a more complete scheme of all processes triggered by the interaction with a VUV photon and their molecular timescales (cf. Fig. \ref{astropah_schematic}). On the very short timescales is ionization.
XUV fs experiments have shown that it occurs on a characteristic time of $40\,\mathrm{fs}$ for small PAHs \cite{Marciniak2015} and this time was found to increase with size. Dissociation starts on longer timescales  \cite{Marciniak2015}. It is found, in the case of PAHs, to be well described by statistical theories. In particular the fitting of breakdown curves obtained in imaging photoelectron photoion coincidence (iPEPICO) spectroscopy at the Swiss Light synchrotron with a model based on the Rice-Ramsperger-Kassel-Marcus (RRKM) theory has been successful to quantify activation energies and dissociation rates by studies in the $1-100\,\mathrm{\mu s}$ range \cite{West2018}. In this time window, dissociation is not in competition with radiative cooling. It is only in the analysis of the experiments in ion traps and storage rings (e.g. in the PIRENEA setup described below) that this competition has to be taken into account considering the involved long timescales.

An interesting result in the last years of the photophysics of PAHs has been the confirmation in the laboratory of recurrent fluorescence as a main radiative cooling mechanism. Recurrent fluorescence, also called Poincar\'{e} fluorescence, has been predicted by L\'{e}ger \textit{et al.} \cite{Leger1988}. Boissel \textit{et al.} \cite{Boissel1997} reported evidence for its role in the cooling of trapped anthracene cations submitted to the radiation of a Xe lamp. The process was confirmed and quantified in the Mini-ring storage ring \cite{Martin2013, Martin2015}.
The use of storage rings has opened the possibility for dynamical studies over the ms window for the study of the fast radiative cooling of energized PAHs. Timescales longer than seconds are now becoming accessible with the new generation of cryogenic rings such as DESIREE at the University of Stockholm or the electrostatic cryogenic storage ring CSR at the Max-Planck Institute for Nuclear Physics and University of Heidelberg. These will be really valuable to address the infrared cooling of PAHs \cite{Stockett2019}.

\section{Methods}
\subsection{Experimental methods}

We have used ion trap experiments to investigate the effect of size on the dissociation of PAHs. More specifically, we discuss here the importance of the C$_2$H$_2$ loss channel and the variation of the H dissociation rate with the PAH size. Two setups have been used: PIRENEA, a dedicated setup for astrochemistry, and the commercial linear ion trap available at the VUV DESIRS beamline. These setups are briefly discussed below.

$\bullet$ The PIRENEA setup for astrochemistry is a cryogenic Fourier transform ion cyclotron resonance mass spectrometer (FTICR-MS) that has been specifically designed to approach the conditions of the interstellar medium in terms of isolation of the trapped species. It is therefore best suited to study the photophysics of PAH ions on long timescales. It is unfortunately not interfaced with a tuneable VUV source, therefore a multiple photon absorption scheme with lower energy photons is used to achieve fragmentation. In this scheme, the photons are absorbed sequentially. Due to fast internal conversion  (timescale typically less than ps \cite{ghanta2011}), the energy absorbed in an excited electronic state is rapidly converted into internal energy of the ground state before  absorption of the next photon. Boissel \textit{et al.} \cite{Boissel1997} have demonstrated that the use of a Xe lamp is convenient to control the heating of the ions and study their dissociation in competition with radiative cooling, which we call the dissociation at threshold. In these conditions, branching ratios between the different fragments can be obtained as illustrated in this article. In these experiments, one can easily follow the relative abundance of the parent and the different fragments as a function of the irradiation time. The fitting of these curves with a kinetic Monte Carlo model can be used to extract a dissociation rate (close to the threshold). The model of Montillaud \textit{et al.} \cite{Montillaud2013} has been built on the analysis of data obtained for the coronene cation, C$_{24}$H$_{12}^+$.

$\bullet$ The VUV DESIRS beamline at the synchrotron SOLEIL is equipped with a Thermo Scientific LTQ XL\texttrademark\ linear ion trap \cite{Milosavljevic2012}. The parent ions are produced from a neutral precursor in a solution by an atmospheric pressure photo-ionization (APPI) source. Inside the trap, the species of interest (given m/z) can be isolated by ejecting other species. The isolated singly charged cations are then irradiated by the tuneable VUV synchrotron radiation and a few hundreds of mass spectra are recorded for each photon energy. The photon rate is in the range of $10^{12}-10^{13}\,\mathrm{photons\,s^{-1}}$. Over the studied $8-20\,\mathrm{eV}$ energy range, the irradiation time (between typically $0.2$ and $0.8\,\mathrm{s}$) and the opening of the exit slit are tuned in order to maximize the signal-to-noise ratio and to minimize multiple photon absorption events, which are therefore rare in our experimental conditions. The resulting photoproducts, \textit{i.e.} fragments (H, H$_2$ and eventually C$_2$H$_2$ loss) and doubly charged cations, are mass-analyzed and action spectra can be built as a function of photon energy. Two campaigns were performed, one on small/medium species up to 24 carbon atoms \cite{Zhen2016, Rodriguezcastillo2018} and one on larger species up to 48 carbon atoms.

\subsection{Numerical simulations}
Molecular dynamics (MD) simulations are performed to study the dissociation of PAH radical cations in their ground electronic states. The electronic structure is described on-the-fly within the self-consistent-charge density functional-based tight binding (SCC-DFTB) scheme \cite{elstner1998self}. This methodology has been found to be promising to study isomerization effects during dissociation and trend in the H/C$_2$H$_2$ branching ratio as a function of internal energy \cite{Simon2017}. 
As the C$_2$H$_2$ channel was found to be overestimated relative to H loss, we 
 adopt here a similar procedure as in ref.\cite{Simon2011} consisting in slightly modifying C-H and H-H interaction potentials by scaling the initial values of the parametrized atomic integrals    $\langle\phi_{\mu}^\mathrm{C,H}\vert\hat{h}[\rho_{0}]\vert\phi_{\nu}^\mathrm{H}\rangle$ and  $\langle\phi_{\mu}^\mathrm{C,H}\vert\phi_{\nu}^\mathrm{H}\rangle$  by 0.95 
($\phi_{\mu,\nu}$ are the atomic orbitals and $\hat{h}[\rho_{0}]$ refers to the monoelectronic Hamiltonian at the reference density). Using this modified potential, we performed hundreds of simulations at the lowest energy values to observe dissociation in a reasonable running time.  

\begin{figure}[H]
\begin{center}
\includegraphics[width=\textwidth]{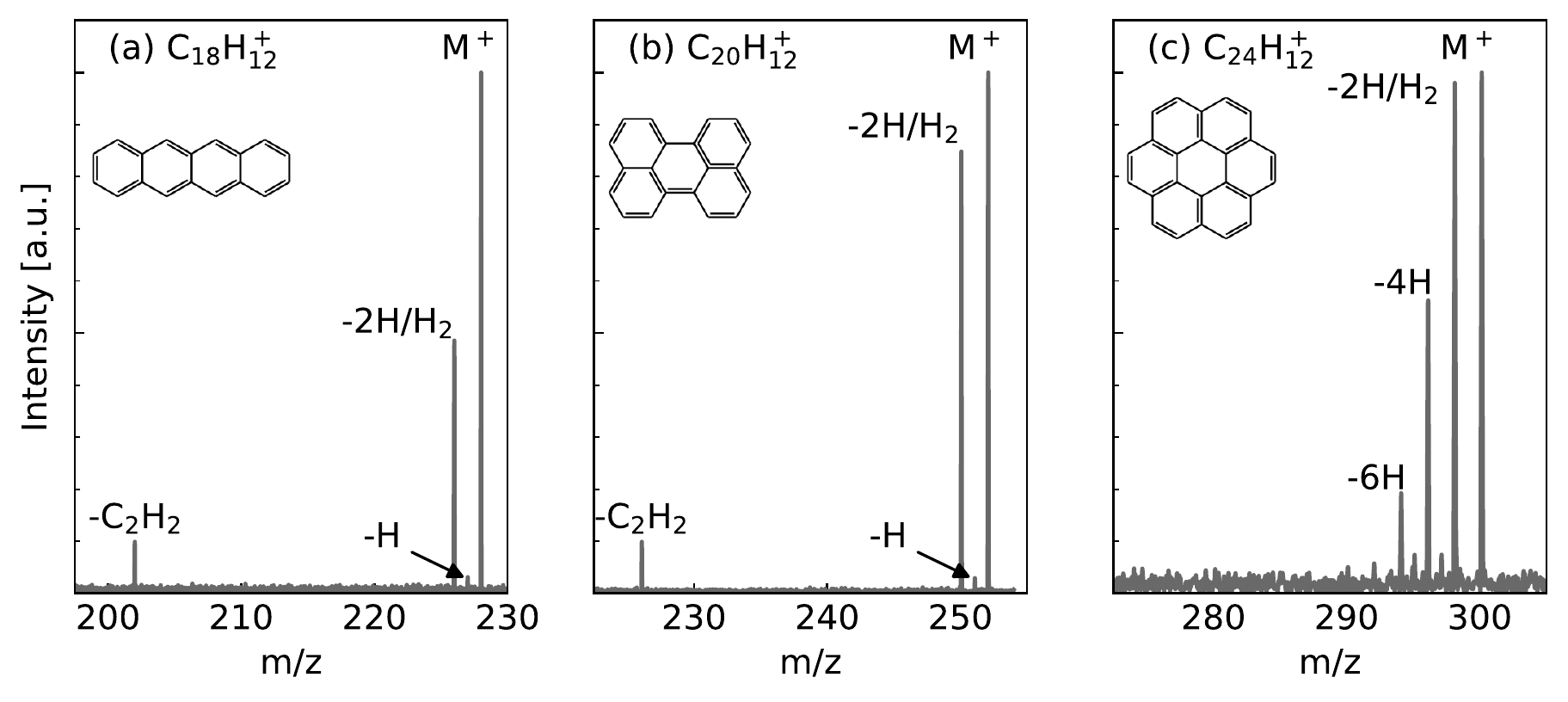}
\vspace{-1cm}
\caption{\label{massspecs}Fragmentation mass spectra of the cations of (a) tetracene, $\mathrm{C}_{18}\mathrm{H}_{12}^+$, (b) perylene, $\mathrm{C}_{20}\mathrm{H}_{12}^+$, and (c) coronene, $\mathrm{C}_{24}\mathrm{H}_{12}^+$. The $^{12}$C isotopomer parent ions were isolated in the cryogenic ion cyclotron resonance cell of PIRENEA and exposed to the UV-visible light of a Xe arc lamp.}
\end{center}
\end{figure}

\section{Results}
\subsection{Loss of C\textsubscript{2}H\textsubscript{2} vs loss of hydrogen}

\begin{wrapfigure}[9]{r}{0.5\linewidth}
\centering
\vspace{-1cm}
\includegraphics[width=\linewidth]{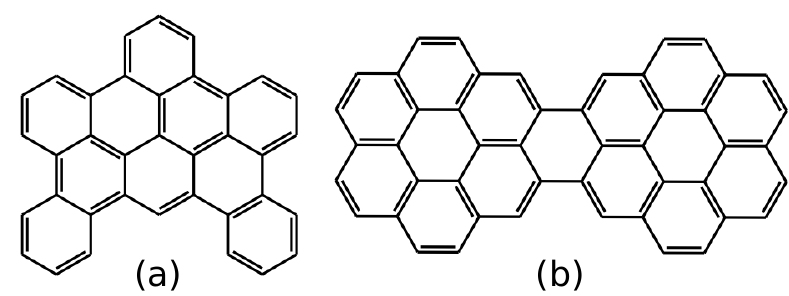}
\vspace{-1cm}
\caption{\label{molstructures}Molecular structures of the studied (a) dibenzophenanthropentaphene, $\mathrm{C}_{36}\mathrm{H}_{18}^+$, and (b) dicoronylene, $\mathrm{C}_{48}\mathrm{H}_{20}^+$, cations.}
\vspace{0.5cm}
\end{wrapfigure}

The fragmentation of PAHs can involve H and H$_2$ loss but also carbonaceous fragments, in general C$_2$H$_2$ but in some cases also C$_4$H$_2$ or CH$_3$ (case of hydrogenated PAHs) \cite{West2018}. The mass spectra measured with the PIRENEA setup following irradiation with the Xe lamp of the isolated $^{12}$C isotopomers of three medium-sized molecules are shown in Fig. \ref{massspecs}. The less compact structure tetracene, C$_{18}$H$_{12}^+$, shows the most intense peak for the C$_2$H$_2$ fragment, followed by perylene, C$_{20}$H$_{12}^+$. The compact ion coronene, C$_{24}$H$_{12}^+$, does not exhibit C$_{2}$H$_{2}$ loss, neither in PIRENEA nor in SOLEIL experiments at all VUV photon energies up to $20\,\mathrm{eV}$ \cite{Zhen2016}. A similar result was obtained for the larger molecules, dicoronylene, C$_{48}$H$_{20}^+$, but also the less compact ion dibenzo[fg,ij]phenanthro[9,10,1,2,3-pqrst]pentaphene, C$_{36}$H$_{18}^+$, whose structures are shown in Fig.~\ref{molstructures}.

\begin{wrapfigure}[22]{l}{0.5\linewidth}
\vspace*{-0.5cm}
\centering
\includegraphics[width=\linewidth]{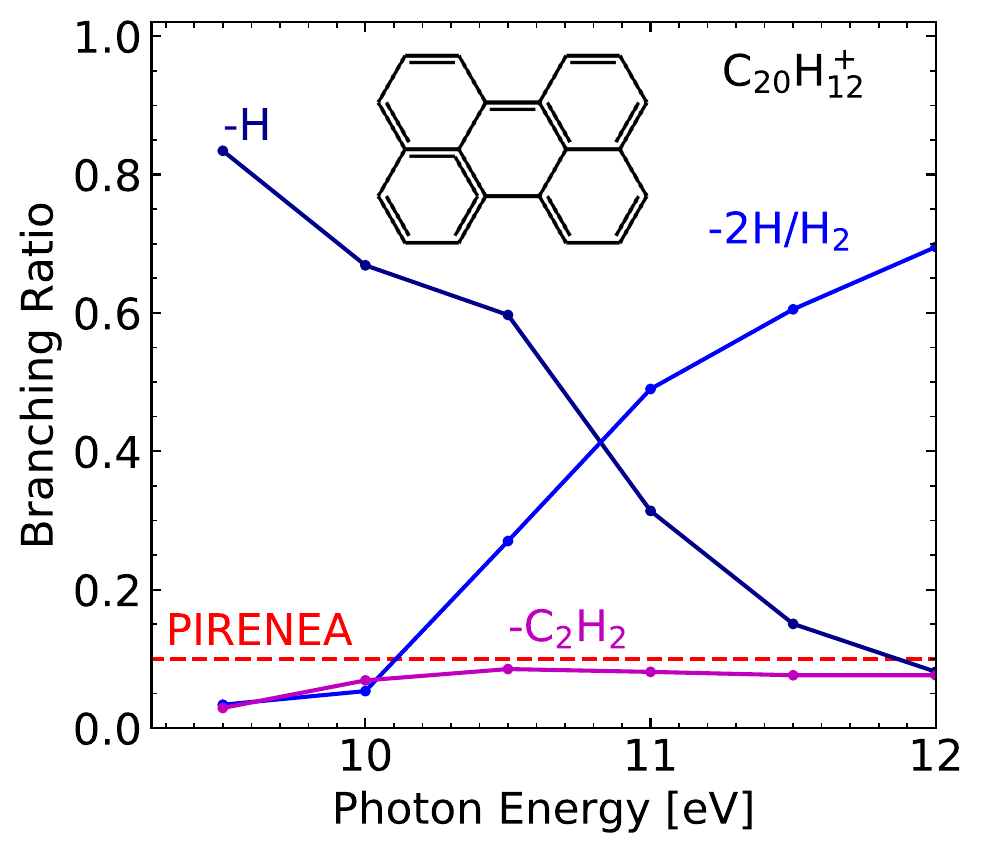}
\vspace*{-1cm}
\caption{\label{branchingratios}Fragmentation branching ratio (BR) for the perylene cation recorded at SOLEIL as a function of the VUV photon energy (first campaign, \cite{Zhen2016}). Also reported is the C$_2$H$_2$/(H+H$_2$) BR derived from the PIRENEA experiments (cf. Fig.~\ref{massspecs}).}
\end{wrapfigure}

 We report in Fig.~\ref{branchingratios} the evolution of the fragmentation branching ratio (BR) with the energy of the absorbed VUV photon for perylene, C$_{20}$H$_{12}^+$, up to $12\,\mathrm{eV}$. In this range, the C$_{2}$H$_{2}$/(H+H$_2$) BR does not vary with energy, whereas the opening of the 2H/H$_2$ channel(s) proceeds. At higher energies, sequential fragmentation is observed and the results are more difficult to interpret. The value of BR recorded at SOLEIL is also consistent with the PIRENEA value, showing that it does not depend on the excitation scheme: multiple UV-visible photon absorption for PIRENEA compared to single VUV photon absorption atå SOLEIL.

The above results appear in line with a statistical fragmentation of the hot ion.
It is therefore of interest to compare them with the results of MD/SCC-DFTB calculations. A total energy of \(24.8\,\mathrm{eV}\) is necessary to observe a reasonable number of fragmentation events for the perylene cation during the simulation time (1620 simulations of 500\,ps).
Averaging over all simulations, losses of H, H$_2$ and C$_2$H$_2$ were observed with BRs of 3.7, 0.6 and 0.7\,\%, respectively. In the case of the coronene cation at the same energy of 0.275\,eV per mode (\textit{i.e.} 28.1 \,eV of internal energy), the H, H$_2$ and C$_2$H$_2$ losses were observed with BRs of 6.2, 0.6 and 0.5\,\%, respectively. The values of 0.08 and 0.07 derived for the C$_{2}$H$_{2}$/H and C$_{2}$H$_{2}$/(H+H$_2$) BRs, are therefore significantly lower than those for the perylene cation at 0.19 and 0.16, respectively. These simulations show the right trends compared to the experiments. Indeed, one should consider that, even with such a large number of simulations, statistics is probably not reached as the number of events at the energy dissociation threshold remains small. MD simulations can also be used to get structural information along the dynamics, which is not accessible in our experiments but can provide interesting insights into dissociation pathways and final products \cite{Jusko2018}. In particular it is interesting to investigate the role of isomerization upon dissociation \cite{Johansson2011}. This is an ongoing work which is not discussed in this article.
\subsection{Dissociation rates}
\label{dissrates}

\begin{wrapfigure}[25]{r}{0.5\linewidth}
\centering
\vspace*{-1cm}
\includegraphics[width=\linewidth]{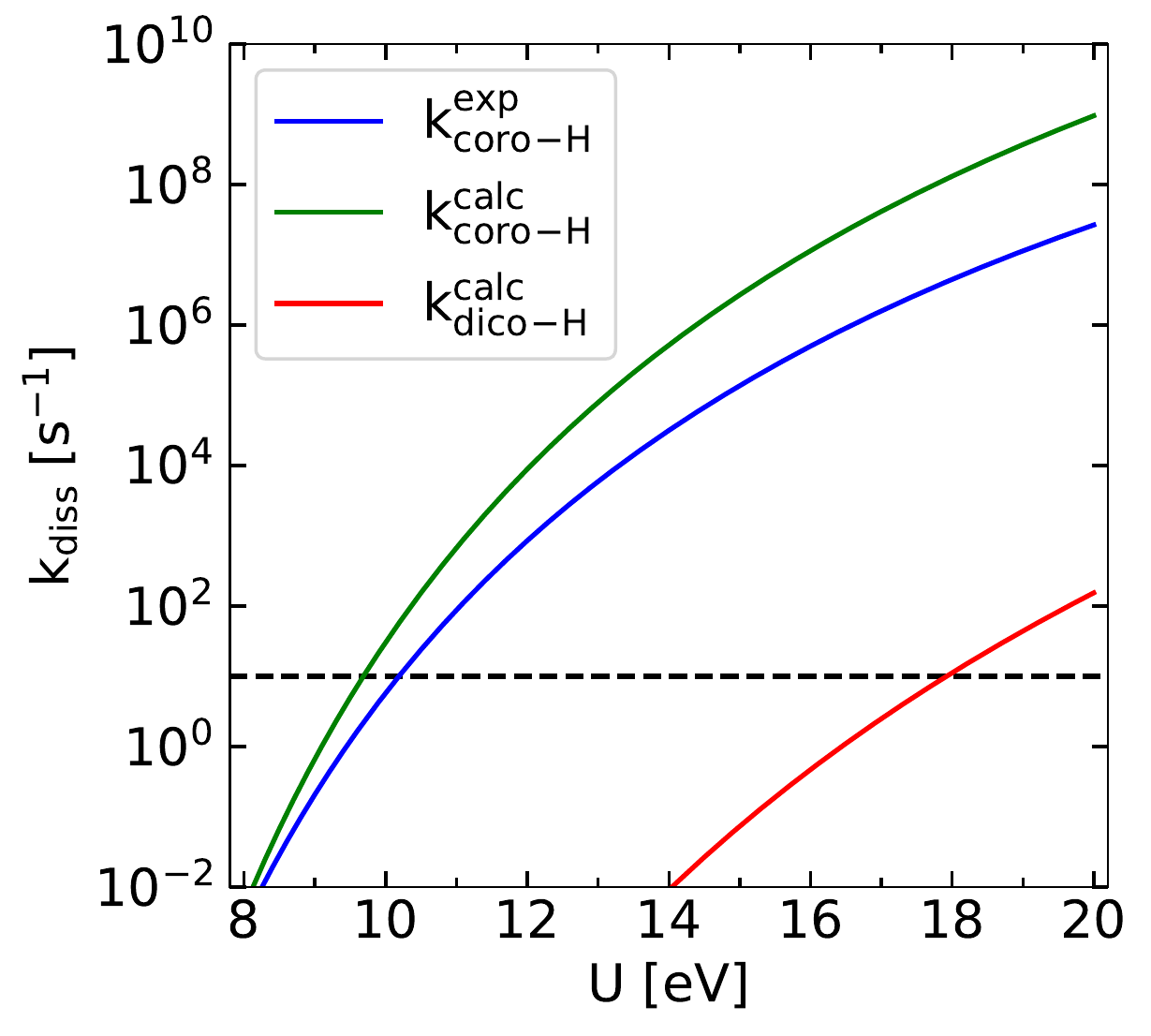}
\vspace*{-1cm}
\caption{\label{kdiss}Dissociation rates of H loss for the coronene (coro) and dicoronylene (dico) cations. For coro, the experimental rate is taken from West \textit{et al.} \cite{West2018} and the calculated one from Montillaud \textit{et al.} \cite{Montillaud2013}. For dico, the calculated rate was obtained from Eq.~\ref{Eq_kdiss} with values of $E_0=4.41\,\mathrm{eV}$ and $A=3.05\cdot10^{15}\,\mathrm{s}^{-1}$ (cf. sect.\,\ref{dissrates}). The horizontal dotted line represents a typical value for the rate of emission of IR photons.}
\end{wrapfigure}

The results of the model by Montillaud \textit{et al.} \cite{Montillaud2013} rely on the dissociation rate for the H loss of the coronene cation that was obtained by fitting the experimental data of the PIRENEA experiment with a kinetic Monte Carlo model and assuming an activation energy of $4.8\,\mathrm{eV}$ based on DFT calculations. This rate is compared in Fig.~\ref{kdiss} with the more recent rate derived from iPEPICO measurements \cite{West2018}. J. R. Barker showed that it is possible to describe the RRKM rate with a simple analytical expression derived from the Laplace transform inversion of the Arrhenius thermal molecular rate \cite{Barker1983}. The two adjustable parameters are the activation energy $E_0$ and a pre-exponential factor $A$ in the equation:
\begin{equation}
k(U) = A \rho(U-E_0)/\rho(U)
\label{Eq_kdiss}
\end{equation}
\noindent where $\rho$ is the density of vibrational states (DoS) of the parent ion and U its internal energy. We computed the DoS using the Beyer \& Swinehart algorithm \cite{Beyer1973} and the list of harmonic vibrational modes listed in the theoretical spectral database of PAHs
\cite{Malloci2007}. We then fitted the dissociation rate of the coronene cation from \cite{West2018} using the mean value of $4.41\,\mathrm{eV}$ reported by the authors for $E_0$ and adjusting the value of $A$ to $3.05~10^{15}\,\mathrm{s}^{-1}$.

It is now of interest to test whether such rates can be extrapolated to large PAH sizes. West \textit{et al.} derived that the value of $E_0=4.41\,eV$ is typical for the loss of the first H in the studied PAHs up to 24 carbon atoms \cite{West2018}. Assuming that $A$ is independent of size, we can then calculate the dissociation rate of the dicoronylene cation, C$_{48}$H$_{20}^+$, using Eq.~\ref{Eq_kdiss}. The result is shown in Fig.~\ref{kdiss} and shows that C$_{48}$H$_{20}^+$ is not expected to lose H below $17-18\,\mathrm{eV}$. This prediction can be tested with our SOLEIL data.

\begin{wrapfigure}[32]{l}{0.5\linewidth}
\vspace{-0.5cm}
\centering
\includegraphics[width=\linewidth]{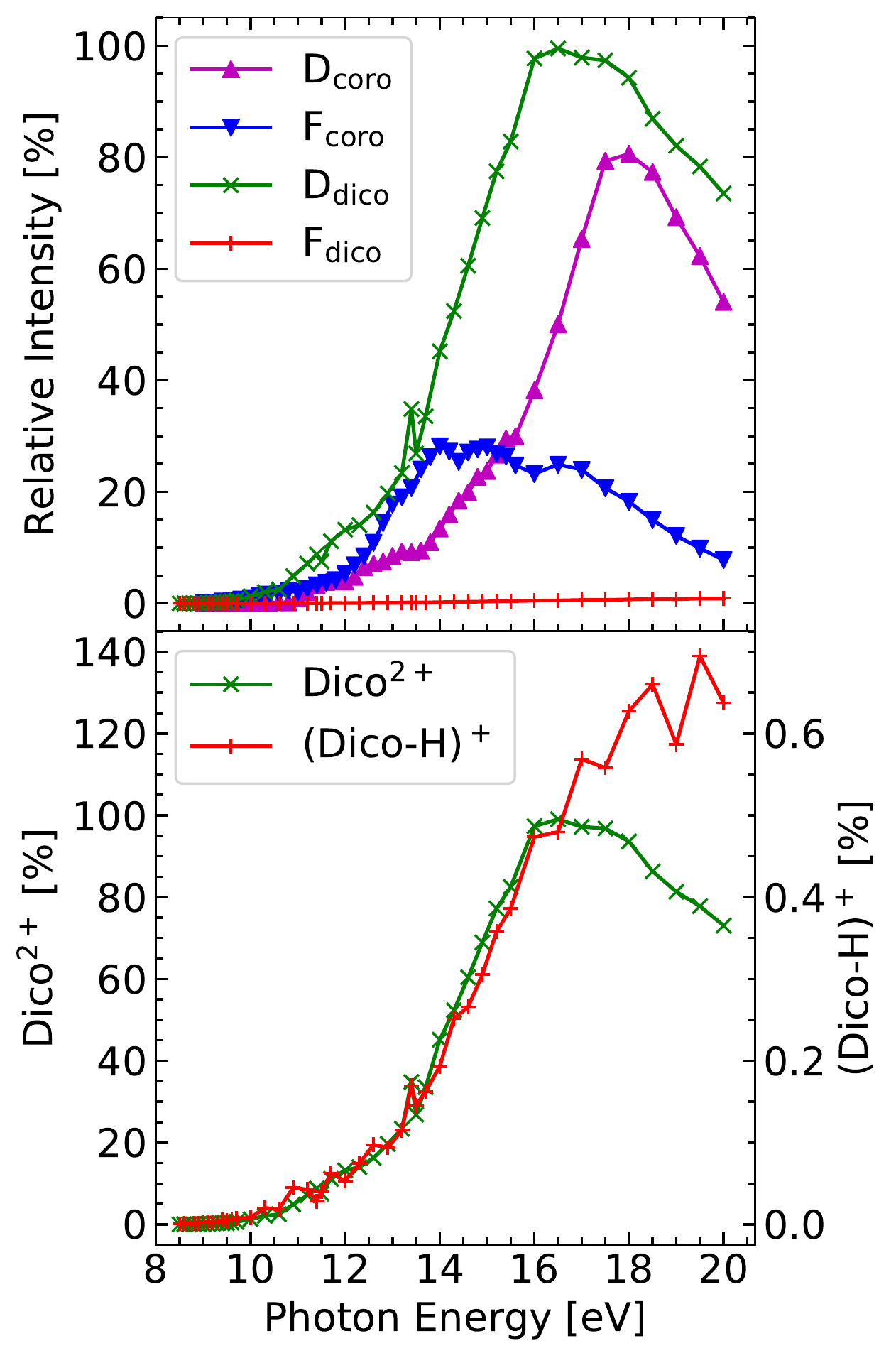}
\vspace*{-1cm}
\caption{\label{Hloss}Upper panel: Dication, D, and fragment, F, photoproducts upon VUV photon absorption of coronene, $\mathrm{C}_{24}\mathrm{H}_{12}^+$ (coro) and dicoronylene, $\mathrm{C}_{48}\mathrm{H}_{20}^+$ (dico). Lower panel: Comparison of the dication and the H loss channel of dicoronylene.}
\end{wrapfigure}

Figure~\ref{Hloss} (upper panel) compares the different photoproducts observed for coronene \cite{Zhen2016} and dicoronylene cations. For the latter, ionization leading to the dication is largely dominating, whereas there is more competition with hydrogen loss in the case of coronene. Zooming on the dicoronylene-H fragment channel (lower panel in Fig.~\ref{Hloss}), one can notice that the signal, although weak, starts to increase at around $12\,\mathrm{eV}$, which strongly disagrees with the predicted dissociation rate (cf. Fig.~\ref{kdiss}). One could think of improper cooling of a small population of ions in the trap, possibly related to the sequential absorption of VUV photons over a short enough period so that these ions had not enough time to cool before the absorption of another VUV photon. However, a closer inspection of our data led us to notice that the fragmentation curve is following very closely the ionization curve until both curves split at $\sim\!17\,\mathrm{eV}$ (lower panel in Fig.~\ref{Hloss}). This suggests that both channels are strictly in competition below $17\,\mathrm{eV}$, which can be rationalized if dissociation is driven by electronic excited states. Thermal dissociation from the hot ground state of the cation would then start when both curves would deviate at $\sim\!17\,\mathrm{eV}$ in reasonable agreement with the calculated dissociation rate.  All this points to the occurrence of non-statistical dissociation through the same intermediate states leading to ionization. This non-statistical dissociation process is different from the one that has been reported for collisions with atoms at center-of-mass energies from a few tens to a few hundreds of eV \cite{Stockett2015}. In this case, specific non-statistical fragments are observed. On the contrary, in our experiments with VUV photons we observe a direct H loss at energies below those expected for statistical fragmentation following internal conversion. Direct dissociation in the excited states in a non-statistical process was reported earlier for large molecules, namely protonated peptides \cite{Gregoire2006} and amino-acids \cite{Gregoire2009} but never for PAHs.

This work combined with a previous study on the hexa-peri-hexabenzocoronene, C$_{42}$H$_{18}^+$, cation \cite{Zhen2015} shows that ionization is by far the dominant process in the processing of large PAHs by VUV photons. In addition, 
our SOLEIL experiments suggest that excited electronic states could play a role in dissociation. So far, only statistical (thermal) dissociation has been considered in astronomical models, which requires energies over $13.6\,\mathrm{eV}$ for large PAHs and therefore rare multiple absorption events \cite{Montillaud2013}. On the other hand, although of very low occurrence relative to ionization, dissociation from excited electronic states can proceed from the absorption of a single VUV photon of energy less than $13.6\,\mathrm{eV}$. It can therefore be very competitive relative to a multiple absorption process.
\section{Conclusion}
The study of the photophysics and stability of PAHs in astrophysical environments has motivated a wealth of studies involving a large variety of experimental setups to address the multiscale dynamics of these systems. Major progresses have been achieved but quantitative studies should now be extended to larger species. We discuss here some results obtained in ion traps for species containing up to $\sim\!50$ carbon atoms and came to the following conclusions. The loss of C$_2$H$_2$ is not expected to occur in the fragmentation of these large PAHs. Ionization largely dominates fragmentation and this competition has to be better evaluated to be used in astrophysical models. In addition, the possibility to dissociate at energies below $13.6\,\mathrm{eV}$, as suggested by our SOLEIL experiments on the dicoronylene cation, has to be further investigated in particular through probing the relaxation of excited electronic states with ultra-fast diagnostics \cite{Marciniak2015}.\\

\ack
The research leading to this result is supported by the European Research Council under the European Union's Seventh Framework Programme ERC-2013-SyG, Grant Agreement no. 610256 NANOCOSMOS. G Wenzel is supported by the H2020-MSCA-ITN-2016 Program EUROPAH, Grant Agreement no. 722346. The computing mesocenter CALMIP (UMS CNRS 3667) is acknowledged for generous allocation of computer resources (p0059).


\section*{References}
\bibliographystyle{iopart-num}
\bibliography{ICPEAC2019}

\end{document}